\documentstyle[12pt]{article}
\newcommand{\lb}[1]{\label{#1}}
\newcommand{\bn}{\begin}
\newcommand{\be}{\begin{equation}}
\newcommand{\ee}{\end{equation}}
\newcommand{\bd}{\begin{displaymath}}
\newcommand{\ed}{\end{displaymath}}
\newcommand{\ci}[1]{\cite{#1}}

\newcommand{\bs}[2]{\section{#1}\label{#2}}
\newcommand{\bea}[1]{\begin{eqnarray}\label{#1}}
\newcommand{\eea}{\end{eqnarray}}
\newcommand{\eref}[1]{{(}\ref{#1}{)}}   
\newcommand{\fref}[1]{\ref{#1}}   
\newcommand{\sref}[1]{\,\ref{#1}}        
\newcommand{\ma}[1]{\mathaccent#1}

\newcommand{\nn}{\nonumber}
\setbox1=\hbox{\LARGE\raise0.5ex\hbox{?`}}
\title{A Quantum Material Model of Static Schwarzschild Black Holes}
\author{S.-T. Sung\thanks{E-mail address: s.t.sung@durham.ac.uk}\\
        Department of Mathematical Sciences\\
        University of Durham, Durham DH1 3LE, U.K.}
\begin{document}
\maketitle
%
%
%
%
%
%
%
\begin{abstract}
A quantum-mechanical prescription of static Einstein field equation is proposed in order to construct the matter-metric eigen-states in the interior of a static Schwarzschild black hole where the signature of space-time is chosen as $(--++)$. The spectrum of the quantum states is identified to be the integral multiples of the surface gravity. A statistical explanation of black hole entropy is given and a quantisation rule for the masses of Schwarzschild black holes is proposed.

\end{abstract}
%
%
%

%
%
%
%
%
%
%
%
\bs{Introduction}{intr}

Schwarzschild solution of the Einstein field equation was discovered eighty years ago. Since then, it inspired many an imagination (see \ci{tho94} for a vivid portrait). In particular, after the introduction of black hole entropy \ci{bek73b}, the formulation of four laws of black hole mechanics \ci{bar73}, the discoveries of Hawking radiation \ci{haw75} and Unruh effect \ci{unr76}, quantum theory in/of curved space-time experienced a boom in the past two decades, as can be seen clearly from the abundant references in SLAC pre-print database \ci{slac}.

The subject which induced this work is the statistical explanation of black hole entropy. As the four laws of black hole mechanics were formulated, the similarity in the mathematical appearance between them and the laws of thermodynamics were regarded as purely superficial \ci{bar73}. Nonetheless, Bekenstein proposed that the area of a black hole should be truly regarded as a measure of the entropy of the black hole \ci{bek73b}. This proposal received a strong support from Hawking's discovery that not only a black hole radiates thermally, but also the temperature of the thermal radiation can be related to the surface gravity, $\kappa$, of the black hole by the relation $T=\kappa/2\pi$ \ci{haw75}. Since then, the similarity between the two sets of laws is regarded as indicating something deeper physically, instead of an accident, and one of the most intensively discussed problems is the statistical origin of black hole entropy.

There have been various proposals about the statistical origin of black hole entropy (see \ci{bek94b} and references therein). We would like to provide an alternative. We attempt to understand this problem from the viewpoint of matter inside a black hole. Hence, we will construct a quantum material model of static Schwarzschild black holes. This could be a bit controversial due to the appearance of singularity predicted by the theory of general relativity \ci{haw73}, which is the gravity theory we adopted in this article. If classical general relativity is still applicable beyond the event horizon, matter will simply fall into the singularity without any sigh. However, the singularity theorems are formulated at classical level. Since a self-consistent quantum gravity is
still beyond our scope, it is unclear how the picture will be changed. 

Nevertheless, it is not too na\"{\i}ve to expect that the quantum effect could turn such a catastrophe into a new chance of surviving---much like what happens inside an atom: If one treats an electron around a proton classically, then the catastrophe of the electron falling into the proton is inevitable. The quantum effect rescues the electrons, and us, from the death penalty of being annihilated. Encouraged by the lesson from atomic physics, we think it makes sense to ask this question: Could this also happen inside a black hole so that the quantum effect can prevent the matter from falling into the singularity?

In order to understand the quantum effect which could happen inside a black hole, we propose a quantum-mechanical prescription of static Einstein field equation which will then be imposed on the matter and the metrics concerned.   

Our approach is inspired by the quantum-mechanical approach to an atom. It will be helpful to appreciate our approach if the reader would like to compare a black hole with an atom frequently. In such a comparison, a static black hole is like an atom heated at a certain temperature. The matter inside a black hole is like the electrons in the atom. The eigen-states inside a black hole and the corresponding energy spectrum of is thus like the eigen-states and energy spectrum of those electrons.

The comparison can also be extended to the dynamic process: a cloud of matter collapsing to form a black hole is like an electron being captured by a proton, black hole radiation is like the emission of photons due to transitions between eigen-states; though, we are not yet able to provide descriptions to these dynamic processes. Although the two systems are totally different dynamically, we hope such a comparison will help the reader to capture the essential conceptual points in our approach. 

Within such a  black hole-atom analogy, our major tasks are the construction of the eigen-states of matter (and the associated metrics) and the identification of the energy spectrum. In section \sref{bhe}, we will foremost give a statistical explanation of black hole entropy. In section \sref{pre}, we will spell out the quantum-mechanical prescription of static Einstein field equation. The quantum material models based on this prescription will be given in section \sref{model1} and \sref{model2}. In concluding section, we will reflect the whole approach again from various point of views and project its future prospects.

%
%
%
%
%
\bs{A statistical explanation of black hole entropy}{bhe}
We at first show, from a phenomenological point of view, how we explain the statistical origin of black hole entropy. If one assumes, as we do, that the statistical aspect of a statistical explanation of black hole entropy can be borrowed from the textbook statistical mechanics, then the two basic ingredients for computing the statistical entropy of a thermal equilibrium system are the spectrum of the states in the system concerned and the statistical distribution law governing these states.

We assume that a static Schwarzschild black hole of mass $M$ is in fact a thermal equilibrium system at temperature $T=\kappa/2\pi=1/8\pi GM$ which is consisted of states whose distributions are governed by Bose-Einstein statistics. Furthermore, we assume that the spectrum of these states is given by $E_j=\kappa j$, $j=1,2,3,...$. Then, the thermal properties of a Bose system can be calculated from the logarithm of the partition function $Z$ \ci{hua87}, (We employ units with $\hbar=c=k_B=1$. The gravitational constant $G$ will be written explicitly.)
\bd
lnZ=-\beta F_b=-N_b\sum^{\infty}_{j=1}ln(1-e^{-\beta E_j})~,
\lb{e1}\ed
where $N_b$ is a normalisation constant and $F_b$ is the Helmholtz free energy. 
We define the following quantities for convenience:
\bea{e2.1'2.2}
n_j=\frac{1}{e^{2\pi
j}-1}~,&b_0=-\sum^\infty_{j=1}ln(\frac{e^{-2\pi j}}{n_j})~,\lb{e2.1}\nonumber\\
b_1=\sum^\infty_{j=1}n_j~,&b_2=\sum^\infty_{j=1}2\pi j n_j~.\lb{e2.2}\nonumber
\eea

Then the relation between the Helmholtz free energy ${F}_b$, internal energy ${U}_b$, and entropy ${S}_b$ is
\be
{U}_b=  -(\partial_{\beta}lnZ)_{E_j,N_b}=T{S}_b+{F}_b= TN_b b_2~,
\lb{e3}\ee
where
\be
{S}_b= \beta^2(\partial_{\beta}{F}_b)_{E_j,N_b}=N_b(b_2+b_0)~.
\lb{e4}\ee
The total number of states, $N$, is
\be
N= N_b\sum^{\infty}_{j=1}n_j=N_b b_1~.
\lb{e5}\ee
Then the entropy per state ${\bar s}_b$ is,
\bd
{\bar s}_b={{S}_b\over N}={b_0+b_2\over b_1}~,
\lb{e6}\ed
which is independent of $M$ due to the specific dependence of $T$ and $E_j$ on $\kappa$. The total derivative of \eref{e3} is
\be
dU_b=TdS_b+S_bdT+dF_b~.
\lb{e7}\ee
We can then identify $\delta Q\equiv TdS_b$ in \eref{e7} with $dM$ in the first law of black hole mechanics. The normalisation constant is then determined to be
\be
N_b={{4\pi GM^2}\over{b_0+b_2}}~.
\lb{e8}\ee
This relation then give us a quantisation rule for the masses of Schwarzschild black holes:
Since the total number of state, $N$, should be a positive integer, consequently, from \eref{e5} we obtain
\bd
M=\sqrt{\frac{b_0+b_2}{4\pi G b_1}N}~.
\lb{e9}\ed
This conforms (up to a prefactor) with various derivations. (See \ci{mak96} and references therein.)

The entropy per state (up to an additional constant) can also be understood in
a different way: Since the relative number of states in state $j$ is ${n}_{j}$,
the entropy per state, ${\bar s}_b^\prime$, is therefore
\bea{e10'}
{\bar s}^{\scriptscriptstyle\prime}_b&=& -{1\over b_1}\sum^{\infty}_{j=1}{n}_{j} ln({{n}_{j}\over b_1})\nonumber\\
&=&{\bar s}_b+ln(b_1)+{1\over b_1}\sum^{\infty}_{j=1}{ln(1-e^{-2\pi j})\over {1-e^{-2\pi j}}}~.\lb{e10}\nonumber
\eea
Within such a picture, the statistical origin of black hole entropy can thus be understood as distributing matter to all possible states. 

We have given a statistical explanation of black hole entropy in the spirit of textbook statistical mechanics under several assumptions. We assumed that the conventional concepts of statistical mechanics can be generalised to a black hole such extreme object because we are not aware of any reason that should motivate us to modify them. Otherwise, one has to invent a version of statistical mechanics to justify one's attempt to explain black hole entropy statistically. Admittedly, the faith can be challenged that the black hole entropy indeed has a statistical origin. For the competitive explanations of (non-statistical) black hole entropy, the reader is referred to literature (for example, see reference \ci{wald94}).

We also assumed that a static black hole can be regarded as a thermal equilibrium system because only which has a well-defined entropy in the statistical mechanics we currently understand. As to the value of the temperature, we are not able to justify that why it should be the temperature of the thermal radiation observed at future infinity if the black hole is radiating, though, in the case we are considering it is not. We {\it assume} the temperature is so.

The rest of this article is then devoted to the construction of the matter-metric eigen-states inside a black hole and the identification of their spectrum as $E_j=\kappa j$. The eigen-states of an electron in an atom are solved from the Schr\"odinger equation. We therefore in next section prescribe the Schr\"odinger equation-like equations suitable for our purpose.  

%
%
%
%
%
\bs{A quantum-mechanical prescription of static Einstein field equation}{pre}
Before we present our approach, we hope the reader bear in mind that our approach is not a quantum field theory in curved space-time though we will borrow some terminology from conventional quantum field theory. In spirit, our approach in the present static case is in fact more quantum-mechanical than quantum-field-theoretical. Tactically, we will try to draw as many lessons and similarities as we can from the historical transition from classical mechanics to quantum mechanics. In order to help understanding our approach, we thus parallel the conventional quantum field theory in curved space-time \ci{bir82} with classical mechanics (or pre-quantum mechanics), then what we are trying to do is developing a kind of quantum mechanics for matter and the space-time geometry it induces.

In quantum field theory in curved space-times \ci{bir82}, the semi-classical Einstein field equation is give by ${\cal R}_{\mu\nu}-1/2g_{\mu\nu}{\cal R}=-8\pi G\langle {\widehat{\cal T}}_{\mu\nu} \rangle$ in which $\cal{R}_{\mu\nu}$ is the Ricci tensor, $\widehat{\cal{T}}_{\mu\nu}$ the energy-momentum tensor, and $G$ the gravitational constant. In words, the geometry is determined by the expectation value of energy-momentum tensor with respect to a certain state. In conventional approach the expectation value is taken with respect to the vacuum state, one then constructs an effective action and renormalises the energy-momentum tensor. In such an approach, the background space-time has to be prescribed beforehand in order to construct the eigen-states of Hamiltonian. Then the field variable can be expanded in terms of these eigen-states. If one wishes to consider the influences of the energy-momentum on the background space-time, one has to resort to considering back-reaction. 

However, we think one of the most important features of the Einstein field equation, which should be respected, is that it has to be solved self-consistently. In the classical level, it means that the Einstein field equation and the Euler-Lagrange equations for the matter which produce the relevant energy-momenta should be used together to determine the geometry and the matter distribution once and for all. This is well-known to be hard to implement.

Furthermore, at quantum-mechanical level, our opinion is that such an attitude should also be preserved so that a matter eigen-state and the associated space-time geometry should be solved together. Different matter eigen-states will induce different space-time geometry according to the Einstein field equation. In other words, a matter eigen-state and the associated space-time geometry together should be regarded as a matter-metric eigen-state. More precisely, the quantum-mechanical, static Einstein field equation can be written as ${\cal R}_{\mu\nu}-1/2g_{\mu\nu}{\cal R}=-8\pi G\langle j\vert\!\!:\!\!{\widehat{\cal T}}_{\mu\nu}\!\!:\!\!\vert j\rangle$ in which $\vert j\rangle$ is the $j$-th eigen-states and $:(~):$ denotes the normal ordering as one used in quantum field theory. Note that the difference between our prescription and that in the conventional quantum field theory in curved space-time lies majorly in the interpretation. However, it is indeed such an interpretative jump that leads a {\it classical} Schr\"odiger equation to a {\it quantum-mechanical} one.

The above quantum-mechanical prescription is understandably irrelevant as the system interested is an astronomical planet system, such as the sun and the earth. Nevertheless, when the system concerned is a black hole, we think it is important to proceed quantum-mechanically as one is required to treat an electron in an atom quantum-mechanically.

Our prescription is not yet precise enough for us to do anything. In the rest of this section, we will confine ourselves to a specific system to see how to implement this prescription practically.

We will consider a spherically symmetric, self-gravitating system of real scalar field. The Euler-Lagrange equation and the (classical) Einstein field equation can be written as (The definitions of the curvature and energy-momentum tensors follow Weinberg's \ci{wei72}. The signature of Lorentzian space-time is $(-+++)$.)
\bea{e11.1'11.4}
\phi^{\prime\prime}+\phi^{\prime}(\frac{1}{r}+\frac{q^\prime}{q}+\frac{h^\prime}{h})+\frac{r^2}{q^2h^2}\partial_t^2\phi=0~,\lb{e11.1}\\
\frac{h^\prime}{h}+8\pi G\left(\frac{1}{2}\frac{r^3}{q^2h^2}(\partial_t\phi)^2-
\frac{r}{2}{\phi^\prime}^2\right)=0~,\lb{e11.2}\\
\frac{h^\prime}{h}-\frac{1}{q}(1-q^\prime)=0~,\lb{e11.3}\\
\frac{h^{\prime\prime}}{h}+\frac{h^\prime}{h}(\frac{3}{2}\frac{q^\prime}{q}-
\frac{1}{2r})+\frac{1}{2}\frac{q^{\prime\prime}}{q}+\frac{1-q^\prime}{rq}
+8\pi G\frac{r^2}{q^2h^2}(\partial_t\phi)^2=0~,\lb{e11.4}
\eea
where a prime denotes the differentiation with respect to $r$. These equations are derived from the action
\bd
I=\frac{1}{2}\int d^4x {\sqrt{\vert g\vert}}(-\frac{{\cal R}}{8\pi G}-\partial_\mu\phi\cdot\partial^\mu\phi)~,
\lb{e12}\ed
and we have written the metric in the standard form \ci{wei72},
\be
ds^2=h^2\frac{q}{r}dt^2+\frac{r}{q}dr^2+r^2d\Omega^2~.
\lb{e13}\ee
We have a spherically symmetric system in mind, so there is no angular dependence in our equations. Since we are giving a quantum-mechanical prescription of {\it static} Einstein field equation, the metric is $t$-independent. There are in fact only three independent equations due to the Bianchi identity. Note that up to now, everything is classical. To implement the quantum-mechanical prescription, we will dress a $\widehat{\mbox{hat}}$ to the field variable ${\phi}$ which will be realised as operator-valued henceforth.

However, before we implement the quantum-mechanical prescription, we have to introduce another important concept in a quantum theory: the probability density function. We expand $\hat\phi$ as (with $E_j>0$)
\bea{e14.1'14.3}
&\hat\phi=\hat\varphi+\hat\varphi^\dagger~,\lb{e14.1}\nn\\
&\hat\varphi=\sum_j \hat\varphi_j=\sum_j \hat a_j e^{-iE_jt}R_j(r)~,\lb{e14.2}\\
&\hat\varphi^\dagger=\sum_j \hat\varphi_j^\dagger=\sum_j \hat a_j^\dagger e^{iE_jt}R_j(r)~,\lb{e14.3}\nn
\eea
where $\hat a_j$ and $\hat a_j^\dagger$ are the annihilation and creation operators for $j$-th eigen-state $\vert j\rangle$ such that $[\hat a_j,\hat a_i^\dagger]=\delta_{ji}$. We can then regard ${\hat J}^t =-ig^{tt}(\hat\varphi^\dagger\cdot\partial_t\hat\varphi-\partial_t\hat\varphi^\dagger\cdot\hat\varphi)$ as the probability density operator so that the function, $\langle j\vert\hat{J}^t\vert j\rangle$, will be identified as the probability density function of matter in a matter-metric eigen-state. The normalisation condition is
\be
\int dr d\Omega \sqrt{\vert g\vert }\hat J^t=-\int dr d\Omega\sqrt{\vert g_s g^{tt}\vert}\langle j\vert {\hat J}_t\vert j\rangle
=N_j~,
\lb{e15}\ee
where $N_j$ is the normalisation of the matter and $g_s$ is the spatial part of the metric in the $j$-th matter-metric eigen-state. This definition is a generalisation of number density operator in the quantum field theory in flat space-time \ci{bir82}.

We would like to remind the reader again that in the present static case our approach is a quantum-mechanical one. The role of $\hat\phi$ is thus more like a superposition of wave functions in quantum mechanics, rather than a field operator in quantum field theory. The terms $exp(-iE_j t)$ and $exp(iE_j t)$ are thus like the $t$-dependent phase term of an energy eigen-state in quantum mechanics. The wave functions in quantum mechanics is realised at operational level, i.e., only the probability density functions (square of the moduli of wave functions) can be associated with experimental outcomes. We also give the probability density function, $\langle j\vert\hat{J}^t\vert j\rangle$, such an operational meaning.

A note is perhaps needed: It is inappropriate to interpret, in either static or dynamic situation, the $\hat\varphi_j$ as propagating on the corresponding space-time geometry, which is solved self-consistently using the quantum-mechanical, static or dynamic Einstein field equation, because the underlying principle of our approach is: There is no prescribed space-time background. The Einstein field equation has to be dealt with in its full value.

In the static case, the situation is just like that in quantum mechanics: we do not say an energy eigen-state of an electron in an atom is propagating in Newtonian space-time. 

A quantum-mechanical, dynamic Einstein field equation can be arrived if the metric is allowed to be $t$-dependent (a prejudicially chosen one). Then the matter and the associated space-time geometry are coexistent at every moment of time $t$. The matter was not born into the space-time. The space-time did not pre-exist the matter. Matter cannot live without a space-time. A space-time stripped off matter does not matter.

Consequently, a question like initial-value problem for the field $\phi$ has to be asked carefully if one would like to adopt the above attitude; one should not try to formulate such a problem in a background space-time, even though that space-time geometry is the solution in a particular matter-metric eigen-state.

A challenging question is: How to interpret the static space-time geometry in a particular matter-metric eigen-state? We interpret it in this manner: Theoretically, it is calculated according the quantum-mechanical, static Einstein field equation.  Experimentally, it is the space-time geometry a particle (i.e., the matter in a matter-metric eigen-state) is experiencing while it is being measured. We need sufficient amount of measurement outcomes to draw a fairly good picture of the probability density function. Similarly, we also need the same amount of measurements to build up the structure of the metric. A single measurement will not tell us anything about the probability density function and the metric. 

We can now implement the quantum-mechanical prescription by replacing $\phi$ in equations \eref{e11.1}--\eref{e11.4} with $\hat\phi$ expanded as in equation \eref{e14.1}. They can be written as follows with the new variables $W=8\pi GR_j^2\langle j\vert\!\! :\! \hat a_j^\dagger {\hat a_j}+\hat a_j {\hat a_j}^\dagger\! :\!\! \vert j\rangle$, $x=E_j r$, and $\bar f$ after taking the expectation value of $\widehat{\cal T}_{\mu\nu}$ with respect to $\vert j\rangle$,
\bea{e16.1'16'4}
\frac{\ma{'177}{W}}{2}+\frac{\ma{'137}{W}}{2}(\dot{\frac{\bar f}{\bar f}}+\frac{1}{x})-
\frac{x^2}{{\bar f}^2}W-\frac{1}{4}\frac{{\ma{'137}{W}}^2}{W}=0~,\lb{e16.1}\\
\frac{\ma{'137}{h}}{h}+
\frac{x^3}{2{\bar f}^2}W-\frac{x}{8}\frac{{\dot W}^2}{W}=0~,\lb{e16.2}\\
{\dot{\frac{\bar f}{\bar f}}}-\frac{1}{\bar q}=0~,\lb{e16.3}\\
\ma{'177}{\bar f}+\ma{'137}{\bar f}(\frac{x^3}{2{\bar f}^2}W-\frac{x}{8}\frac{\ma{'137}{W^2}}{W})=0~,\lb{e16.4}
\eea
where 
\bd
\bar f{=}{\bar q} h{=}E_j qh{=}E_jf~,
\lb{e17}\ed
and a dot denotes differentiation with respect to $x$.

Before we turn to next section, where we will solve the matter-metric eigen-states inside a black hole using the prescription just given, there is a remark for the operation of normal ordering. We will interpret this operation in a line similar to that in quantum field theory. Therefore, we are only interested in the difference of energies, though the definition of energy has always been an intriguing issue when the theory of general relativity is involved. It is unclear to us if it is possible to formulate a quantum theory without ever mentioning things like Hamiltonian or energy. (Even though this can be achieved theoretically, it is unclear if experimenters will be happy with it.) Our present understanding, experimentally and theoretically, about the universe depends on the concept of energy so much that we will try to conform ourselves with this fact at his moment.

To avoid the introduction of operators, one can consider the prescription that the $\phi$'s in the probability density operator and energy-momentum tensor are replaced with $\varphi_j$ (or $\varphi_j^\dagger$ so that the energy-momentum is real), instead of $\hat\phi$. Then the equations \eref{e16.1}--\eref{e16.4} can be recovered with a proper re-definition of $W$. The detailed formalism will depend on how one introduces dynamics, particularly the interactions, into the theory.

In the present static situation, the detailed dynamic prescription is unimportant. We thus go on to construct a quantum material model of black holes. 

%
%
%
%
%
%
\bs{A quantum material model of Schwarzschild black holes}{model1}
The prescription given in previous section is intended to be independent of the systems we are dealing with, though, practically, not all systems need to be treated in such a manner. In this section, we will construct a model of static Schwarzschild black holes based on the quantum-mechanical, static Einstein field equation.

In order to motivate our approach, we at first consider the conventional quantum field theory on the background of the interior of a Schwarzschild black hole. We write the background metric in the standard form as in equation \eref{e13} and decompose an eigen-state of the massless real scalar field as (with $E_j>0$)
\be
\phi_j=e^{-iE_jt}R_j(r)+ c.c~,
\lb{e17.5}\ee
like that in equation \eref{e14.2}. The Euler-Lagrange equation \eref{e11.1} is then reduced to
\be
R^{\prime\prime}+R^{\prime}({1\over r}+{q^\prime\over q}+{h^\prime\over
h})={E^2r^2\over q^2h^2}R~.
\lb{e18}\ee

It can be easily checked that $q=r-r_s$ ($r_s=2GM$ in which $M$ is the mass of the black hole) and $h=constant$ is a solution of the vacuum Einstein field equation. The choice of $h=i$ then corresponds to the well-known
Schwarzschild solution.  From equation \eref{e18} we clearly see that if
we choose $h=i$, then the wave function oscillates as $(r_s-r)^{\pm ir_sE}+c.c.$ as $r$ tends to
$r_s$ from inside. However, if we choose $h=1$ (the signature of space-time is therefore $(--++)$), then we could have bound states within the region $0<r<r_s$ which asymptotically behave as $(r_s-r)^{+r_sE}$ as $r$ tends to $r_s$. Though these wave functions then diverge logarithmically near the origin $r=0$, they are normalisable. But, how to identify the spectrum?

Similar to the Euclidean Schwarzschild solution \ci{gib77b}, the Kleinian solution (i.e., with signature $(--++)$) can also be derived from the Lorentzian one through analytic continuation. By introducing Kruskal co-ordinates, we can write the metric of the Lorentzian Schwarzschild solution as \ci{wald84}
\bd
d s^2=e^{-r/r_s}\frac{4{r_s}^3}{r}(-dT^2+dX^2)+r^2d\Omega~,
\lb{e19}\ed
where $t$ and $r$ are related to $T$ and $X$ by relations
\bea{e20.1'20.2}
e^{t/r_s}&=&\frac{T+X}{X-T}~,\lb{e20.1}\nonumber\\
(\frac{r}{r_s}-1)e^{r/r_s}&=&X^2-T^2.\lb{e20.2}\nonumber
\eea
If we analytically continue $T$ to $-iT$ and $t$ to $-it$, we then arrive at the Euclidean Schwarzschild solution with the constraint $r>r_s$. However, instead of $T$, we can analytically continue $X$ to $iX$. Combined with continuing $t$ to $-it$, we arrive at the Kleinian Schwarzschild solution which is confined within the interior of the black hole with the restriction that $-1<-(T^2+X^2)<0$ in which the end points $-1$ and $0$ corresponding to $r=0$ and $r=r_s$, respectively. As in the Euclidean solution, $t$ is also required to be periodic with period $\beta=4\pi r_s$ in the Kleinian case to avoid conical singularity. Back to equation \eref{e17.5}, we are thus constrained to choose the spectrum as
\be
E_j=\frac{j}{2r_s}=\kappa j~,\hspace{3 em}j=1,2,3,...~,
\lb{e21}\ee
where $\kappa$ ($={1}/{2r_s}$) is the surface gravity. Note that this spectrum is the one we used in section \sref{bhe} to calculate the statistical entropy of a black hole.

On this ground, we thus match an exterior Lorentzian Schwarzschild solution to an interior Kleinian Schwarzschild solution which corresponds to the choice of real $h$ (we choose $h>0$). This definitely raises alarm questioning how we cope with the two time-like directions in a Kleinian space-time; even more, we compactified one of them so that it is periodic. A particle whose trajectory is required only to be time-like could travel around by moving along the non-$t$ time-like direction with $t$ co-ordinate frozen. If it does travel along the compactified time-like direction, a closed time-like curve could form. More basically, what is the concept of a particle of which our understandings have always been associated with the Lorentzian space-time?  We therefore refer to a state on a Kleinian space-time as a generalised {\it state}; moreover, as one is deemed to arrive at incorrect conclusions if one treats an electron around a proton classically, we think only a genuine quantum theory (in the sense given in section \sref{pre}) makes sense in a Kleinian space-time. 

In the static situation (with respect to the prejudicially chosen $t$), we can bypass undesired physical consequences by requiring a Kleinian space-time be confined within a region which is classically inaccessible to us, say the interior of a black hole. (However, this by no means prevents us from learning something from or influenced by the quantum effects produced by a Kleinian space-time, say through scattering.) Note that the energy (i.e., the $E_j$) eigen-states can only be constructed in a static space-time. Furthermore, the static matter-metric eigen-states are realised at operational level, whether we have abilities to measure it practically or no. Hence, the two time-like directions and the compactification of one of them will not cause concerns.

In the dynamic situation, it is unclear to us at this moment if a Kleinian space-time could survive, or, what kind of physical constraints should be imposed on it. Let us come back to the example of an atom and consider the dynamic process of an electron being captured by a proton. Before an electron settles down to form an atom, the atom is not yet an atom (in the static sense). In the conventional approach of scattering, such a static state is the final state.  Within the whole (theoretically) dynamic process, the atom does not exist at all. If one adopts such an attitude to the formation of a black hole, the static black hole then corresponds to the final static state; a Kleinian space-time consequently should not appear in the dynamic process theoretically. In other words, a Kleinian space-time could only appear in the static situation as described in previous paragraph. We thus escape any embarrassment caused by a Kleinian space-time by simply wiping the static black holes off the edge of the universe. Even though Kleinian space-times eventually appear dynamically, it is very likely that they can only serve as the intermediate states, as the virtual particles in quantum field theory, in a process like collision of black holes. Therefore, by requiring a Kleinian space-time be either statically confined or dynamically transitory, the reader can still live a happy life.

Lorentzian space-time is one of the building foundations of quantum field theory. Nonetheless, with non-locality (Einstein-Podolski-Rosen paradox, for example), virtual particles such multitude possibilities in mind and on the ground that, classically, we have no access to the events inside a black hole, we will keep an open mind on the issue of the signature change.

A technical problem we have to face at this moment is the possibility of matching an exterior Lorentzian Schwarzschild solution to an interior Kleinian one. In other words, there are certain junction (matching) conditions at the boundary of the two space-times to be satisfied. Since the space-time is static, it suffices to consider the junction conditions on a hypersurface of $t=constant$. We will adopt the junction conditions given in reference \ci{mit73} by requiring the induce metric and the extrinsic curvature of the 2-D boundary in the 3-D hypersurface be continuous because there is no surface layer. These conditions are satisfied because the extrinsic curvature is zero and the induced metric is $r^2d\Omega^2$. These junction conditions will be part of the criteria of choosing the boundary conditions of matter-metric eigen-states inside the black hole.

Though we have recovered the spectrum \eref{e21} desired, there is one grave unsatisfactoriness: Since we intend to interpret the quantum field as the constituent components of the black hole, the metric should not be the vacuum solution of the Einstein field equation. We thus have to take the energy-momentum tensor of the quantum field into account, i.e., we need to implement the quantum mechanical, static Einstein field equation \eref{e16.1}--\eref{e16.4}.

We then regard the metric variable $h$, $q$, and the $W$ as unknown variables, with proper boundary conditions at $x_i\sim x_s=E_j r_s$, we can solve them numerically \ci{sung97u}.

We chose the following boundary condition,
\bea{e22.1'22.2}
\bar q\sim -(x_s-x)+{\bar q}_{j+1}(x_s-x)^{j+1}~,\lb{e22.1}\nonumber\\
h\sim 1+h_j(x_s-x)^j~,\hspace{1 em}W\sim {w_j}(x_s-x)^j~.\lb{e22.2}\nonumber
\eea
We then find the following self-consistent conditions
\bea{e23.1'23.2}
x_s=\frac{j}{2}~,\hspace{2 em}j=1,2,3,...~,\lb{e23.1}\nonumber\\
{\bar q}_{j+1}=h_j=\frac{-j}{j+1} \frac{w_j}{2}~,\lb{e23.2}\nonumber
\eea
where $w_j$ is a free parameter and will be determined by the normalisation condition of $W$.

Note that we have chosen $j$ as positive integers which is the consequence of the choice $E_j=\kappa j$. Though we have no compulsive reason to make such a choice, this is a nature one if we parametrise the metric as
\be
ds^2=f^\prime\frac{f}{r}(dt^2+\frac{\rho^2}{f_0^2}d\rho^2)+r^2d\Omega^2~,
\lb{e24}\ee
where $f_0=\rho-\rho_s$, $\rho_s=r_s$, and $\rho$ is implicitly defined by the equation
\be
\frac{d\rho}{dr}=\frac{f_0}{\rho}\frac{r}{f}~.
\lb{e25}\ee
By comparing the co-ordinates $t$ and $\rho$ in equation \eref{e24} with the co-ordinates $t$ and $r$ in a Kleinian Schwarzschild solution, it is thus nature to endow $t$ with the character of an angular co-ordinate. Above choice of spectrum is thus demanded for topological reason. We will call the solution of $\bar f$ and $W$ corresponding to $j$ the $j$-th eigen-state. Note that we do not regard the exterior region of the black hole as part of those eigen-states. 

From equations \eref{e16.1}--\eref{e16.4} we can discover the asymptotic behaviour at $x\sim 0$ of the $j$-th eigen-state:
$h\sim x^{a_j}$, ${\bar q}\sim x^{-a_j}$, and $W\sim {2a_j}
ln^2(x)$ in which
$a_j$ is positive. $W$ is then always normalisable (see equation \eref{e15}). The integral of the expectation value of $_t$-$^t$ component of the energy-momentum tensor is
\bea{e26'}
\cal{E}&\stackrel{def}{=}&\;\; \int drd\Omega {\sqrt{\vert{g}\vert}}\langle j\vert\!\!:\!{\hat {\cal T}_t}^t\!:\!\!\vert j\rangle\nn\\
&=&\frac{-1}{2G E_j}\int^{j/2}_0 dx{\vert \bar{f}\vert}(\frac{1}{2}\frac{x^3}{\bar{f}^2}W-\frac{x}{8}\frac{\dot{W}^2}{W})\nn\\
&=&\frac{1}{2G E_j}\int^{j/2}_{0} dx {\vert{\bar f}\vert}\frac{\dot h}{h}~.\lb{e26}
\eea
It diverges logarithmically.

Using $h$-$q$ parametrisation in equation \eref{e13}, we found that it is possible to obtain sensible solutions for the fully implemented quantum-mechanical, static Einstein field equation. However, There are two drawbacks in the model based on the $h$-$q$ parametrisation. The first, since the analytical solution is unavailable, it is a bit tricky to integrate the equation \eref{e25}. Particularly, we are interested in knowing the range of co-ordinate $\rho$. The second, the ${\cal E}$ in equation \eref{e26} is divergent. It should be interesting to construct another model to bypass these two problems.

%
%
%
%
%
%
\bs{An alternative model}{model2}

We consider a model based on a different parametrisation of the metric. We parametrise the metric as
\be
ds^2=\frac{\eta}{r}(dt^2+\frac{\rho^2}{f_0^2}d\rho^2)+r^2d\Omega^2~,
\lb{e27}\ee
with $\eta$ and $r$ being regarded as unknown variables. Then it is a straightforward exercise to write down the Euler-Lagrange and Einstein field equations. With proper initial condition at $y_i\sim y_s=E_j\rho_s$, we will show numerically that the boundary condition at $y_s$ and the asymptotic behaviour at $y\sim 0$ given below can be linked together (see figures \fref{f1}--\fref{f6}) \ci{matlab}. We will give the relevant equations directly,
\bea{e28.1'e28.3}
\frac{\ddot{W}}{2}+\frac{\ma{'137}{W}}{2}(\frac{\ma{'137}{\bar f_0}}{\bar f_0}-\frac{1}{y}+2\frac{\ma{'137}{x}}{x})-\frac{y^2}{{\bar f}_0^2}W-\frac{1}{4}\frac{{\ma{'137}{W}}^2}{W}=0~,\lb{e28.1}\\
\frac{1}{2}\frac{(x^2)^{\ma{'177}{}}}{x^2}+\frac{1}{2}\frac{(x^2)^{\mathaccent'137{}}}{x^2}(\frac{\dot{\bar{f}_0}}{\bar{f}_0}-\frac{1}{y})-\frac{{\bar\eta}y^2}{x^3{\bar f_0}^2}=0~,\lb{e28.2}\\
{\ddot{\bar\sigma}}+\dot{\bar\sigma}(\frac{\dot{\bar{f}_0}}{\bar{f}_0}-\frac{1}{y})+\frac{\bar\eta y^2}{x^3{\bar f_0}^2}+\frac{y^2}{{\bar f_0}^2}W+\frac{1}{4}\frac{{\dot W}^2}{W}=0~,\lb{e28.3}
\eea
where $y=E_j \rho$, $\bar\sigma=ln(-\bar\eta)$, ${\bar f}_0=y-y_s$, $x=E_j r$, and a dot denotes differentiation with respect to $y$.

We consider the following boundary conditions at $y_{s}$,
\bea{e29.1'e29.2}
 \bar \eta\sim -(y_{s}-y)+{\bar \eta}_{j+1}(y_{s}-y)^{j+1}~,\hspace{2 em}\lb{e29.1}\nn\\
W\sim w_j(y_{s}-y)^j~,\hspace{1 em}x\sim y+x_{j+1}(y_{s}-y)^{j+1}~,\lb{e29.2}
\eea
where $y_{s}=j/2,j=1,2,3...$, then $x_{j+1}$ and $\bar \eta_{j+1}$ are determined by $w_j$ from the following relations,
\be
\bar \eta_{j+1}=\frac{w_j}{2}~,
\hspace{ 1 em}x_{j+1}=\frac{\bar\eta_{j+1}}{(j+1)^2}~.
\lb{e30}\ee
In deriving above relations, the spectrum $E_j=\kappa j$ ($j=1,2,3,...,$) has been chosen to conform with the interpretation that the $t$ co-ordinate in the metric \eref{e27} has the character of an angular co-ordinate so that $t$ is periodic with a period of $2\pi/\kappa$. The choice of $w_j$ will be determined by the normalisation condition of $W$.

Asymptotically near $y=0$, $W$, $x$, and $\bar\eta$ behave as
\be
W\sim w_0+w_2y^2~,\hspace{1 em} x\sim x_0+x_2y^2~,\hspace{1 em}\bar \eta\sim  \bar\eta_0+{\bar \eta}_{2}y^{2}~,
\lb{e31}\ee
where $w_i$, $x_i$, and $\bar\eta_i$ ($i=1,2$) are constants. The numerical results suggest that $x_0\not= 0$ (see figure \fref{fx}), in contrast to the model based on the $h$-$q$ parametrisation in which the range of $x$ is $0<x<x_s=E_jr_s$ . The origin of the difference lies on the boundary conditions. For the $h$-$q$ parametrisation \eref{e13}, it can be derived, using equation \eref{e25} by requiring self-consistence, that $x=y+o\left((y_s-y)^{j+2}\right)$, in contrast to equation \eref{e29.2}.

With the help of the boundary condition \eref{e29.2} and the asymptotic behaviour \eref{e31}, it is seen that the integral of the expectation value of $_t$-$^t$ component of the energy-momentum tensor for any eigen-state is finite because
\bea{e32'}
\cal{E}&=&4\pi\int^{\rho_{s}}_0 d\rho\sqrt{\vert g\vert}\langle j\vert\!\!:\!{\widehat{\cal T}_t}^t\!:\!\!\vert j\rangle\nonumber\\
 &=&\frac{-1}{2GE_j}\int^{j/2}_0 dy\frac{yx^2}{\vert \bar f_0\vert}\left( \frac{1}{2}W-\frac{1}{8}\frac{{\bar f}_0^2}{y^2}\frac{{\dot W}^2}{W}\right)\nn\\
&\stackrel{def}{=}&\frac{1}{2GE_j}\int^{j/2}_0 dy {\cal E}_j(y)~.\lb{e32}
\eea

The numerical value of $w_j$ are determined by the normalisation condition \eref{e15} with $N_j$ yet to be specified. Recall that in our statistical explanation of black hole entropy in section \sref{bhe}, the probability of finding matter in the $j$-th eigen-state is $Nn_j$. Therefore, we should set $N_j=Nn_j$. With the help of equations \eref{e5}, \eref{e8}, and the definition of $\kappa$, equation \eref{e15} can be reduced to
\be
\int^{j/2}_{0} dy \frac{yx^2}{\vert\bar f_0\vert}W\stackrel{def}{=}\int^{j/2}_0dy {\cal W}_j(y) =\frac{\pi}{2}\frac{j^2n_j}{b_0+b_2}~.
\lb{e33}\ee
Note that the above expression is independent of the masses of black holes.

%
%
%
%
%
\bs{Conclusions}{con}

As a theory, our approach is only a starter. Objections can be easily raised and a lot of questions are awaiting to be addressed. Nonetheless, we are trying to see things from a point of view that is different from the one adopted in conventional quantum field theory which we believe has its limitation. The same idea can also be applied to charged black holes \ci{sung97b}.

In order to understand the statistical origin (in the sense of the textbook statistical mechanics) of black hole entropy, we felt obliged to put matter into a black hole. In order to construct matter bound states, we rotated the signature of the interior space-time of a black hole to Kleinian type. We also gave a quantum-mechanical prescription of static Einstein field equation as the building foundation of the matter-metric eigen-states. None of these is easy to justify by itself standing along. However, things began to make sense as they were combined to form a logical argument. (We hope our logical presentation is clean enough for the reader to follow.)

The earlier literature of which we know discussing physics in a four-dimensional Kleinian space-time is quite rare. The one which is potentially relevant to our approach is given in reference \ci{alt94}. Because of the unusual causal structure, it is reasonable to expect some dramatic physical phenomena emerging in a Kleinian space-time, as reported in reference \ci{alt94}. Perhaps it is indeed this unusualness that keeps people away from such a space-time. However, if a Kleinian space-time is confined within a region which is inaccessible to us classically, we need stronger reason to rule it out. Admittedly, it is also difficult to justify one's claims about the physics in a Kleinian space-time if such a region is not accessible to us.

Nonetheless, when a black hole is involved, we think we have at least one reason to be optimistic: If a black hole was formed from collapsing matter which eventually settled down to a static, or quasi-static state, the dynamic process of signature change should have left some imprints on the exterior region of the black hole, for example, the modification of the thermal black hole radiation. Such traces should depend on the signature type and the detailed dynamics of the underlying theory: As reported in reference \ci{alt94}, when particles encounter a boundary, the absorption and reflection properties of the boundary depends on the signature type of the other region.

Though a fully implemented quantum mechanical, dynamic Einstein field equation is hard to manage, things are not so gloomy. Some numerical simulations indicate that the power of modern computers is capable of handling some specific situations \ci{cho93}. As an alternative, one can adopt the hybrid of semi-classical plus quantum-mechanical approach. Hawking radiation is an example showing us that even a semi-classical approach by itself is enough to teach us a lot. Though, the mechanism of signature change is still awaiting for exploring.

When dynamics is involved, usually we can divide, by hand, the whole theoretical structure into two parts: the initial and final states, and the propagators. The Schr\"odinger equation is usually used to obtain the initial and final states (in most of the cases they are energy eigen-sates). However, the dynamics described by the propagators is handled better by Feynman's path integral approach \ci{fey65}. Therefore, it will be helpful to understand the dynamic Einstein field equation from the point of view of path integral. Such an approach has been applied to the theory of general relativity in the program of Euclidean quantum gravity \ci{gah93}. The gravitational degrees of freedom is notoriously difficult to handle. However, under imposing certain symmetries and cut-off in the numbers of degrees of freedom, the program of quantum cosmology proliferated \ci{har83} (see also \ci{fan87}). In particular, it will be interesting to understand how a black hole could emit, say photons, when there are transitions happening between those eigen-states inside a black hole. 

As mentioned in the text, the concept of energy, hence time, in the theory of general relativity is not so clear. We have chosen a time variable by hand. We do not think this is a drawback of our approach; nonetheless, it does reflect one of the most basic questions we have to face as dynamics is brought in. The so-called {\it problem of time} has been re-appearing again and again in different contexts. We are not able to review the various opinions at this moment. However, a relevant question that can be asked immediately about our approach is: We have got a finite energy ${\cal E}$ (see \eref{e32}), what are we going to do with it? Can we associate the black hole mass $M$ to any quantities calculated locally? We also encounter another (thermodynamic) energy in the calculation of statistical entropy. What are the relations between all these energy terms? A complete {\it physical} picture of black holes can emerge only after these questions have been answered satisfactory. 

Finally, we leave several remarks.

If the reader is more interested in arriving at non-singular space-time (by non-singular we mean $g_{\mu\nu}$ and $g^{\mu\nu}$ are finite) at the centre of the black hole, the following parametrisation should be used,
\bd
ds^2=\frac{\eta}{r}(dt^2+\frac{r^2}{(\rho-\rho_s)^2}d\rho^2)+r^2d\Omega^2~.
\lb{e34}\ed
Our numerical solution suggest that using the boundary condition \eref{e29.2}, \eref{e30}, $\eta\sim \eta_0+\eta_1\rho+\eta_2\rho^2$ and $r\sim r_0+r_1\rho+r_2\rho^2$ as $\rho\sim 0$ with $\eta_0, r_0\not= 0$ \ci{sung97u}. (The equations are slightly different using this new parametrisation, however the boundary condition \eref{e29.2}, \eref{e30} is still applicable if one imposes $E_j=\kappa j$.) 

In our model, the matter-metric eigen-states are confined within the black hole totally. We do not include the exterior space-time geometry of the black hole as part of an eigen-state. The reason for such an interpretation is: we feel more comfortable living in a purely classical space-time. Since all eigen-states in our model have the same exterior region, it will not cause any interpretative or technical changes. However, if the transition between different matter-metric eigen-sates is allowed to happen, then it seems necessary to include the exterior region as part of a matter-metric eigen-state because as the contents of a black hole changes, its radius should change accordingly. 

Furthermore, it is not impossible to modify the interpretation and the normalisation condition so that the boundary of a static black hole is allowed to be fussy. In other words, different eigen-states have different boundary (i.e., the radius are different). Nonetheless, we wonder if such a situation is technically, or conceptually, or interpretatively preferable.

We rotated the signature of the interior of a black hole to Kleinian type in order to confine the matter totally inside a black hole. However, as a criterion of a bound state, it is enough to require the wave function decaying exponentially. It will be relevant to understand by allowing the wave function exponentially decaying outside a black hole, if it is possible to construct a model without ever changing the signature of space-time.

We have given two models based on different parametrisations and boundary conditions. Obviously, there is no reason to rule out the possibility of constructing further models based on other parametrisations and boundary conditions, or totally different approaches. As far as our approach is concerned, it will be important for us, as a guidance, to choose a proper parametrisation and boundary conditions if further physical criteria for choosing them can be given. The parametrisation in equation \eref{e27} seems to be distinctive in the sense that the $\eta/r$ and $r^2$ can be regarded as conformal factors of a Kleinian Schwarzschild black hole on the two topological sectors, $R^2$ and $S^2$, respectively. If this suggest anything deeper is still awaiting for investigation. 

The final remark is that, if we introduce the simplest coupling between the gravity and the scalar field, then the action is 
\bd
I=\frac{1}{2}\int d^4 x \sqrt{\vert g\vert}(-\frac{{\cal R}}{8\pi G}-\partial_\mu\phi\cdot\partial^\mu\phi-\xi {\cal R}\phi^2)~,  
\lb{e35}\ed
where $\xi$ is a coupling constant such that $\xi=0$ and $\xi=-\frac{1}{6}$ correspond to the minimally and conformally coupled cases, respectively. The Euler-Lagrange equation and one component of the Einstein field equation are (with $\eta$-$r$ parametrisation)
\bea{e36.1'36.2}
\frac{W^{\prime\prime}}{2}+\frac{W^\prime}{2}(\frac{f_0^\prime}{f_0}-\frac{1}{\rho}+2\frac{r^\prime}{r})-\frac{1}{4}\frac{{W^\prime}^2}{W}-\frac{\rho^2}{f_0^2}E_j^2 W-\xi\frac{\eta\rho^2}{rf_0^2}W{\cal R}=0~,\lb{e36.1}\nn\\
(1+\xi W)(\frac{\eta^\prime}{\eta}\frac{r^\prime}{r}-\frac{\eta\rho^2}{r^3f_0^2}) +\frac{1}{2}(\frac{\rho^2}{f_0^2}E_j^2 W-\frac{1}{4}\frac{{W^\prime}^2}{W}) \hspace{7.5 em}\nn\\
-\frac{\xi}{1+6\xi}\frac{\eta\rho^2}{rf_0^2}{\cal R}-\frac{\xi}{2}W^{\prime\prime}-2\xi(\frac{r^2f_0}{\eta\rho})^\prime\frac{\eta\rho}{r^2f_0}W=0~,\lb{e36.2}\nn
\eea
where a prime denotes the differentiation with respect to $\rho$, and $\xi=-1/6$, ${\cal R}=0$ for the conformally coupled case, $\xi\not=0,-1/6$,
\bd
{\cal R}=\frac{-(1+6\xi)}{1+\xi(1+6\xi) W}(\frac{r}{\eta}E^2 W+\frac{1}{4}\frac{rf_0^2}{\eta\rho^2}\frac{{W^\prime}^2}{W})
\lb{e37}\ed
for the general case. Given the ansatz \eref{e29.2} at $\rho\sim \rho_s$, it is found that no self-consistent solution exists due to the appearance of $W^{\prime\prime}$ in the Einstein field equation. Admittedly, our statement does not serve as a proof. Nonetheless, it could hint that the minimally coupled case is prestigious.

{\it Acknowledgements.} The author would like to thank an anonymous referee for helpful comments and bringing reference \ci{alt94} to my attention.
%
%
%
%
%
%
%
%
%
%
%
%
%
%
%
\bn{thebibliography}{99}

\bibitem{tho94}K.S. Thorne, {\it Black Hole and Time Warps: Einstein's Outrageous Legacy} (Picador, London, 1994) 

\bibitem{bek73b}J.D. Bekenstein, Phys. Rev. D{\bf 7} (1973) 2333

\bibitem{bar73}
J.M. Bardeen, B. Carter, and S.W. Hawking, Commun. Math. Phys. {\bf 31} (1973) 161

\bibitem{haw75}S.W. Hawking, Commun. Math. Phys. {\bf 43} (1975) 199

\bibitem{unr76}W.G. Unruh, Phys. Rev. D{\bf 14} (1976) 870

\bibitem{slac}Internet: http://www-spires.slac.standford.edu/find/hep

\bibitem{bek94b}J.D. Bekenstein, {\it Do We Understand Black Hole Entropy?} Talk given at 7th Marcel Grossmann Meeting on General Relativity (MG 7), Stanford, CA, 24-30 July 1994. (gr-qc/9409015)

\bibitem{haw73}S.W. Hawking and G.F.R. Ellis, {\it The Large Scale Structure of Space-time} (Cambridge University Press, Cambridge, 1973)

\bibitem{hua87}K. Huang, {\it Statistical Mechanics}, 2nd edition (John Wiley \& Sons, New York, 1987)

\bibitem{mak96}J. M\"akel\"a, {\it Black Hole Spectrum: Continuous or Discrete?} (gr-qc/9609001)

\bibitem{wald94}R.M. Wald, Phys. Rev. D{\bf 48} (1993) 3427 (gr-qc/9305022)

\bibitem{bir82}
N.D. Birrell and P.C.W. Davies, {\it Quantum Fields in Curved Space} (CUP, Cambridge, 1982)

\bibitem{wei72}S. Weinberg, {\it Gravitation and Cosmology} (Wiley, New York, 1972)

\bibitem{gib77b}G.W. Gibbons and S.W. Hawking, Phys. Rev. D{\bf 15} (1977) 2752

\bibitem{wald84}R.M. Wald, {\it General Relativity} (The University of Chicago Press, Chicago, 1984)

\bibitem{mit73}C.W. Misner, K.S. Thorne, and J.A. Wheeler, {\it Gravitation} (W.H. Freeman and Company, San Francisco, 1973)

\bibitem{sung97u}S.-T. Sung, Ph.D. Thesis (in progress)

\bibitem{matlab} Matlab, Version 4.2C (The MathWorks, Inc.)

\bibitem{sung97b}S.-T. Sung, {\it A Quantum Material Model of Static, Non-rotating, Charged Black Holes} (in preparation)

\bibitem{alt94}L. Alty, Class. Quant. Grav. {\bf 11} (1994) 2523

\bibitem{cho93}M.W. Choptuik, Phys. Rev. Lett. {\bf 70} (1993) 9

\bibitem{fey65}R.P. Feynman and A.R. Hibbs, {\it Quantum Mechanics and Path Integrals} (McGraw-Hill Book Company, New York, 1965)

\bibitem{gah93}G.W. Gibbons and S.W. Hawking, {\it Euclidean Quantum Gravity} (World Scientific, Singapore, 1993)

\bibitem{har83}J.B. Hartle and S.W. Hawking, Phys. Rev. D{\bf 28} (1983) 2960

\bibitem{fan87}L.-Z. Fang and R. Ruffini, eds., {\it Quantum Cosmology} (World Scientific, Singapore, 1987)

\end{thebibliography}
\pagebreak
%
%
%
%
%
%
%
%
%
%
%
%
\bn{figure}\section*{Figure Captions}
\vspace{0 in}
\caption{Figure of $W(y)$. We present the numerical solutions corresponding to states $j=1,2,3$ which are shown as solid, dashed, and dotted lines in various figures. The initial condition is set at $y_i=y_s-10^{-10}$ with $w_1=11.75$, $w_2=0.037$, and $w_3=5.5\times 10^{-5}$. In order to provide a better view of various curves, we introduce a magnifying factor corresponding to $j$-th eigen-state, $m_j$, so that the curves shown are multiplied by a factor $m_j$. In present figure, $(m_1,m_2,m_3)=(1,20,10^4)$.}\lb{f1}
%
\vspace{0 in}
\caption{Figure of $\dot{W}(y)$. $(m_1,m_2,m_3)=(1,20,10^{5})$.}\lb{f2}
%
\vspace{0 in}
\caption{Figure of $\eta (y)$. $(m_1,m_2,m_3)=(1,1,1)$.}\lb{f3}
%
\vspace{0 in}
\caption{Figure of $\dot\eta (y)$. $(m_1,m_2,m_3)=(1,1,1)$.}\lb{f4}
%
\vspace{0 in}
\caption{Figure of $x(y)$. $(m_1,m_2,m_3)=(1,1,1)$.}\lb{fx}\lb{f5}
%
\vspace{0 in}
\caption{Figure of $\dot{x}(y)$. $(m_1,m_2,m_3)=(1,1,1)$.}\lb{f6}
%
\vspace{0 in}
\caption{Figure of ${\cal W}_j(y)$  defined in equation \eref{e33}. $(m_1,m_2,m_3)=(1,10^{2},10^{4})$.}\lb{f7}
%
\vspace{0 in}
\caption{Figure of ${\cal E}_{j}(y)$ defined in equation \eref{e32}. $(m_1,m_2,m_3)=(1,10,10^{4})$.}\lb{f8}
\end{figure}

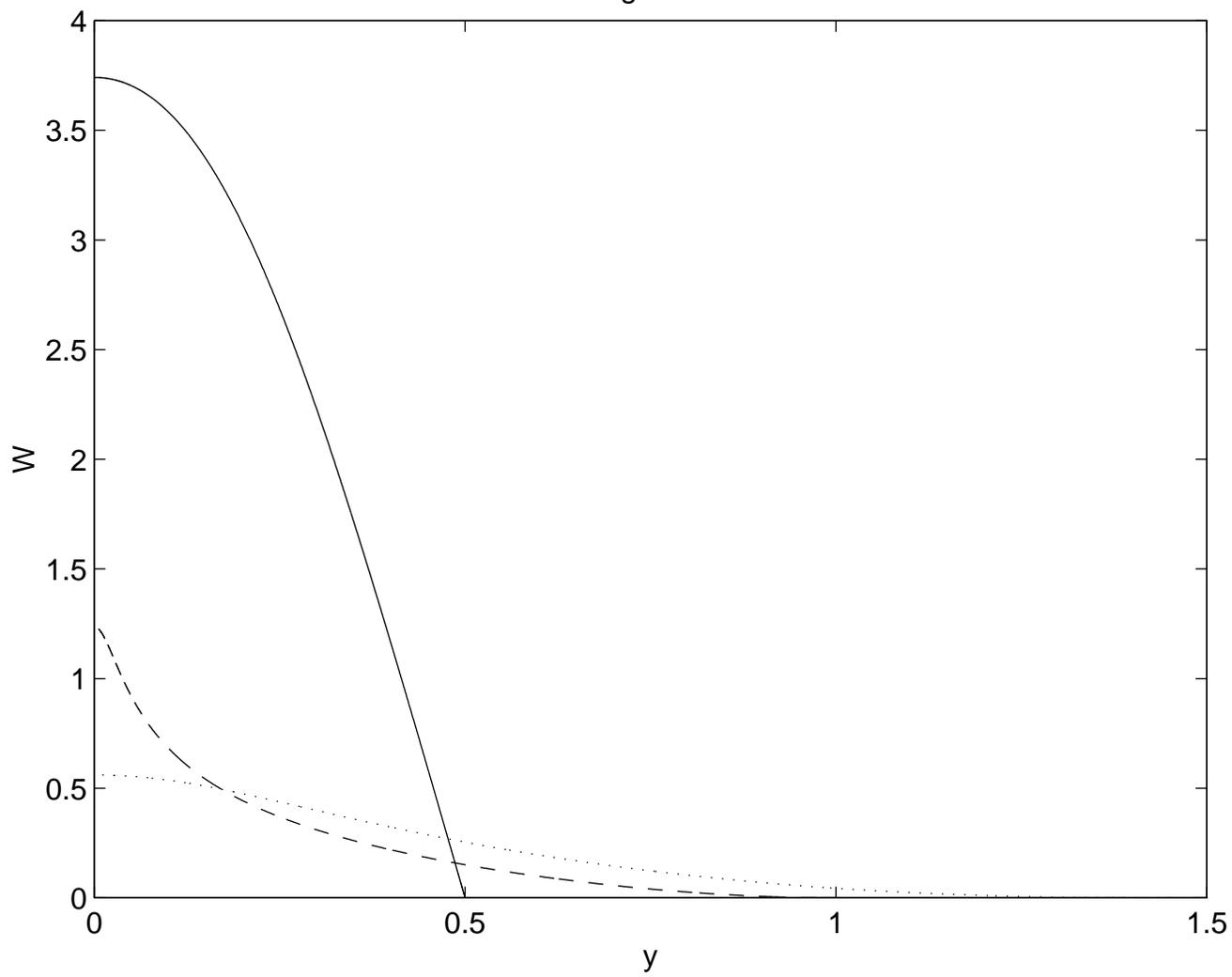
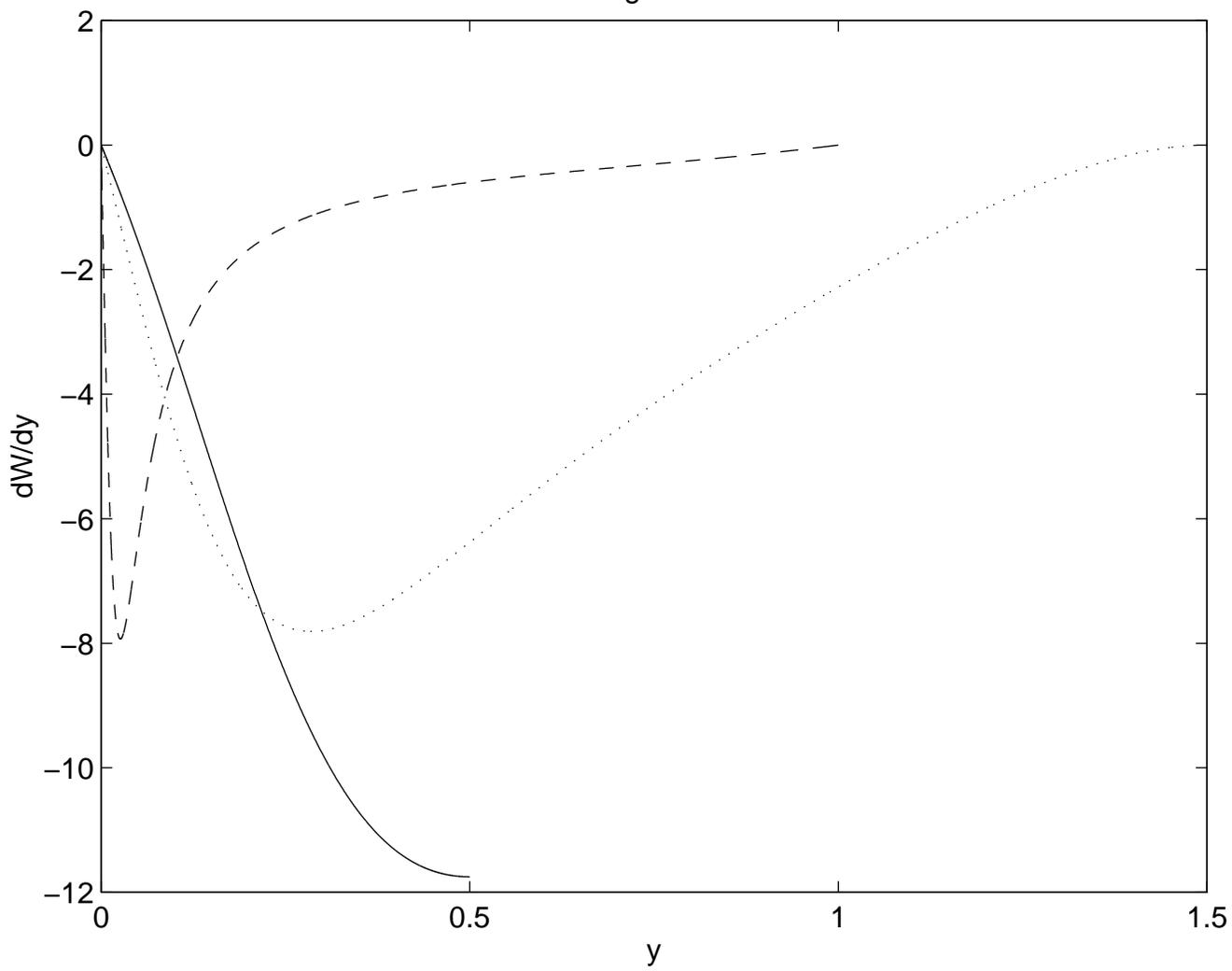
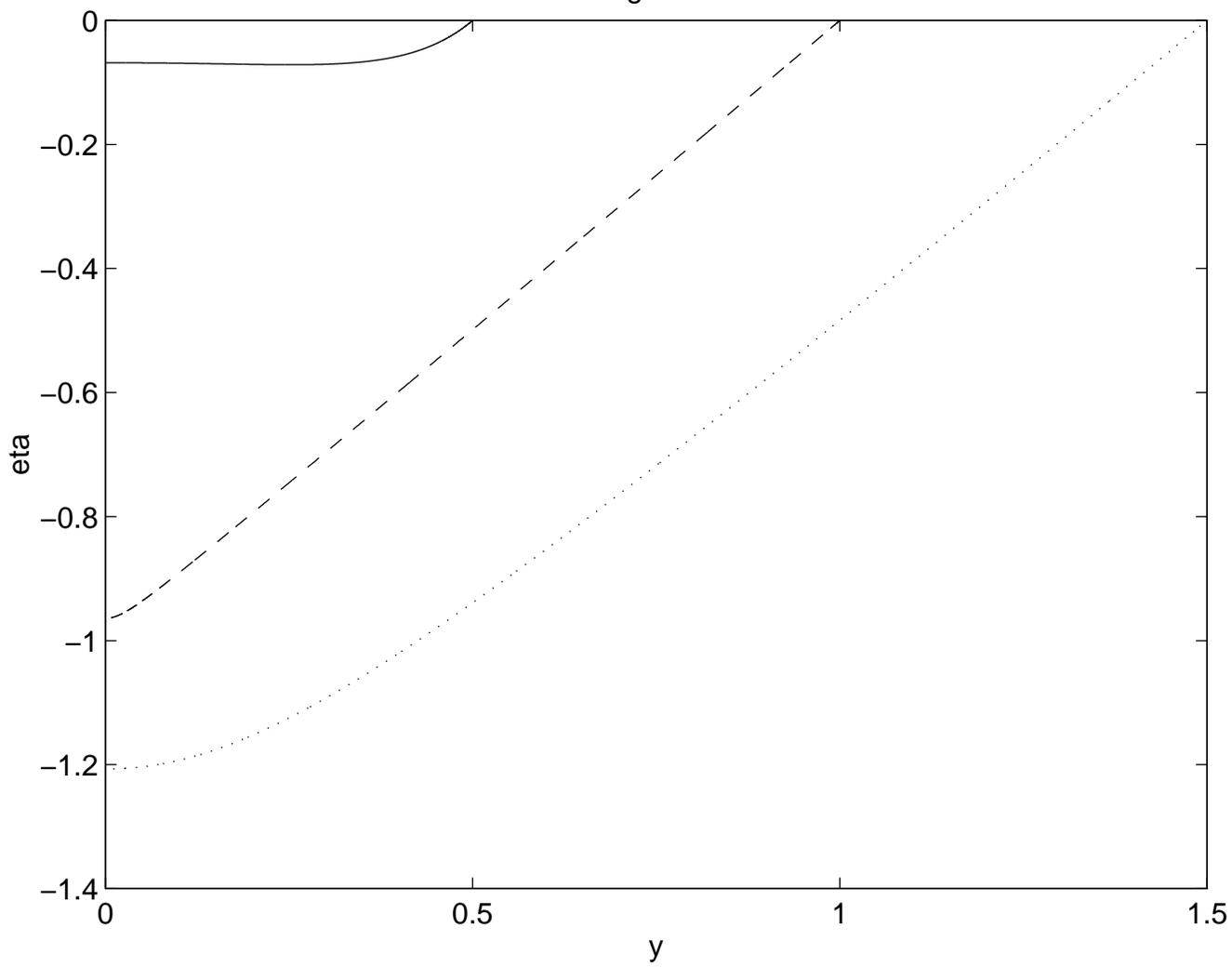
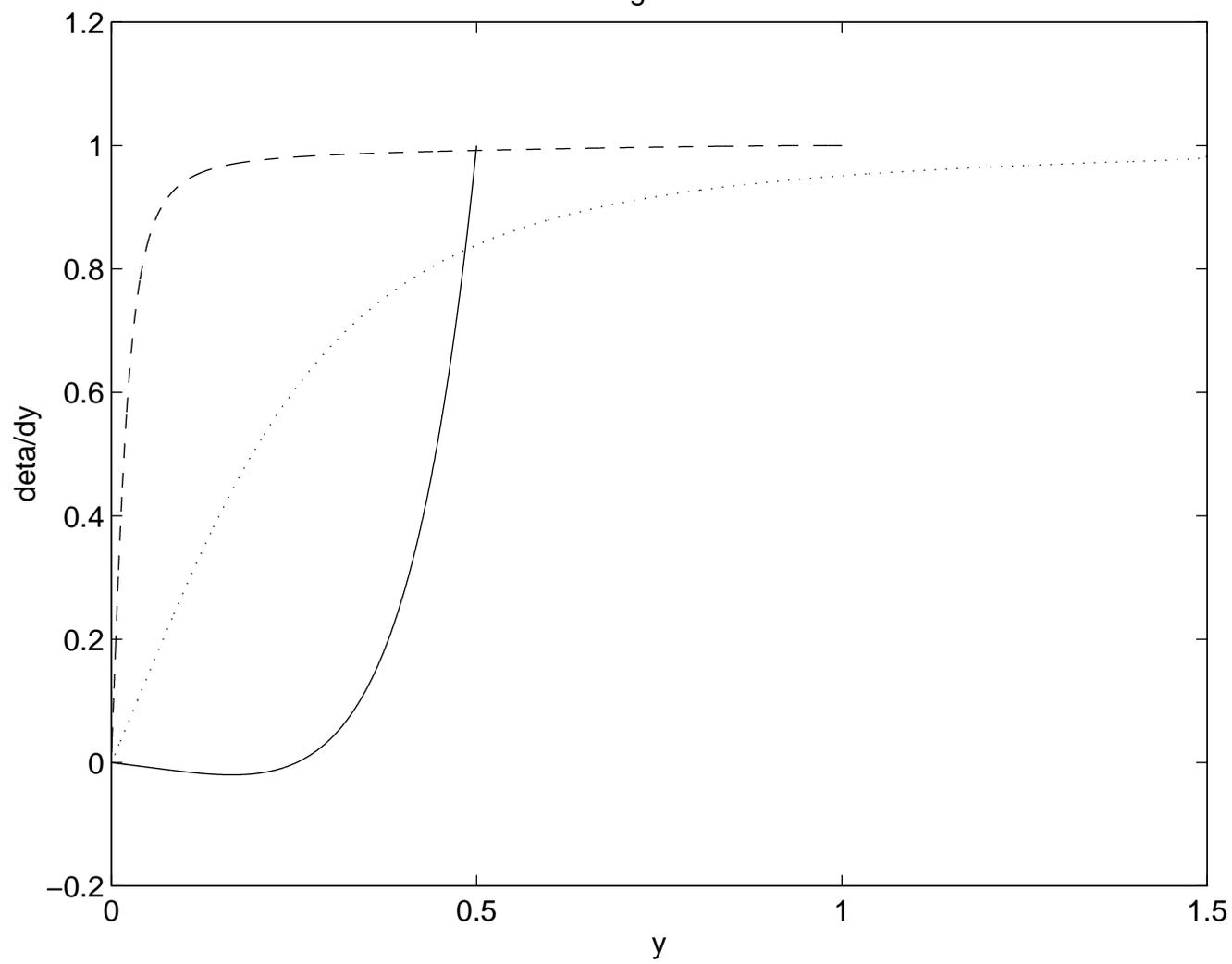
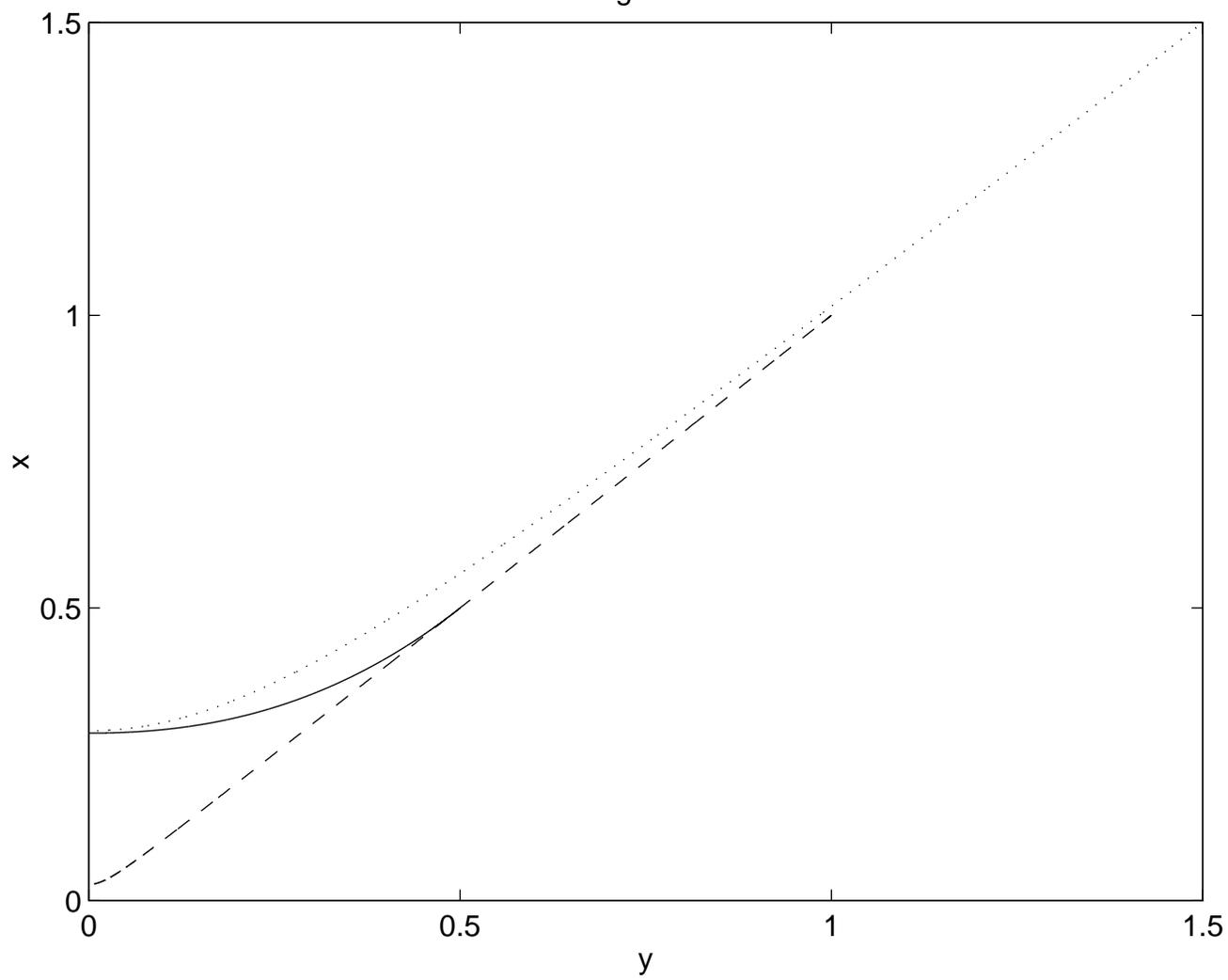
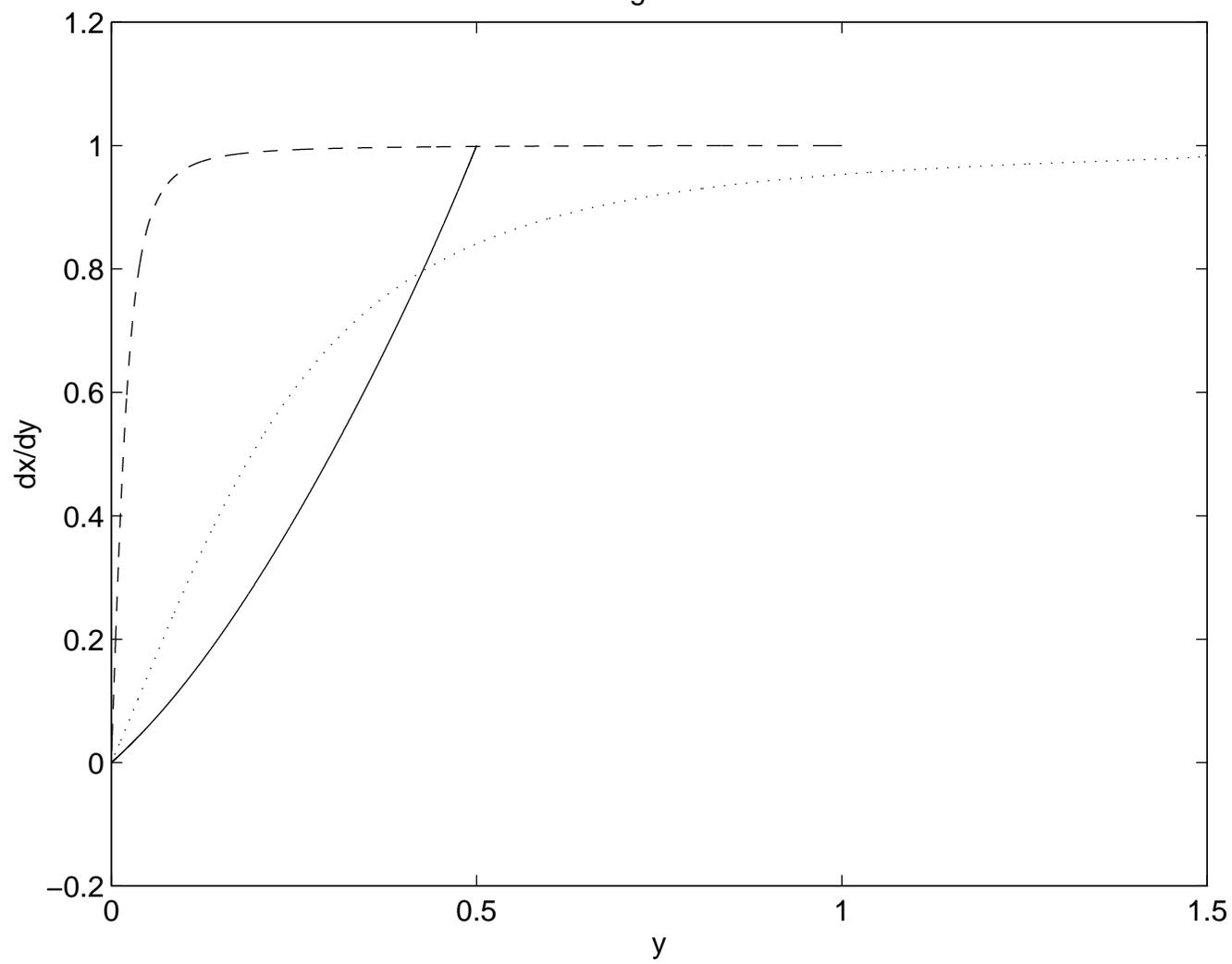
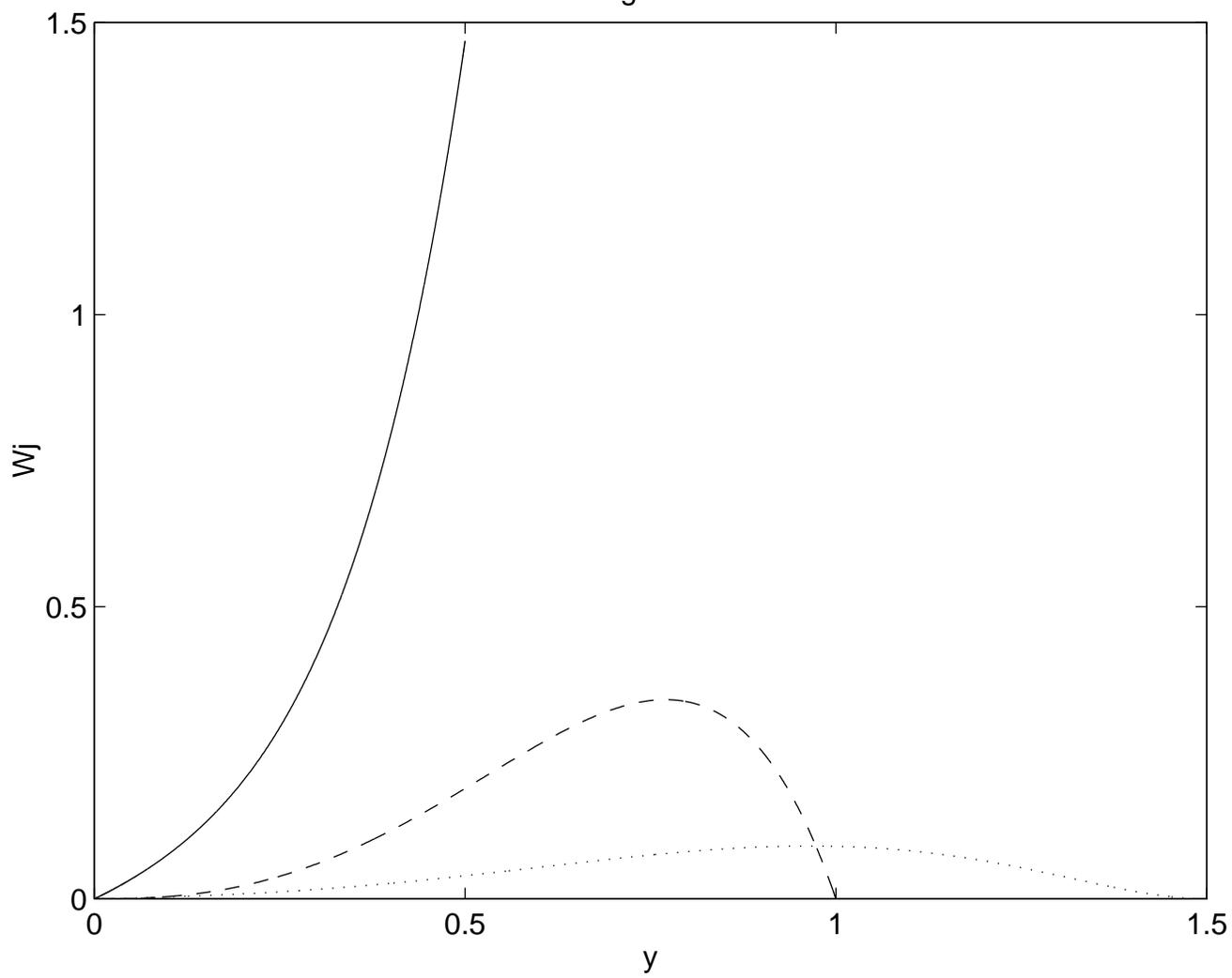
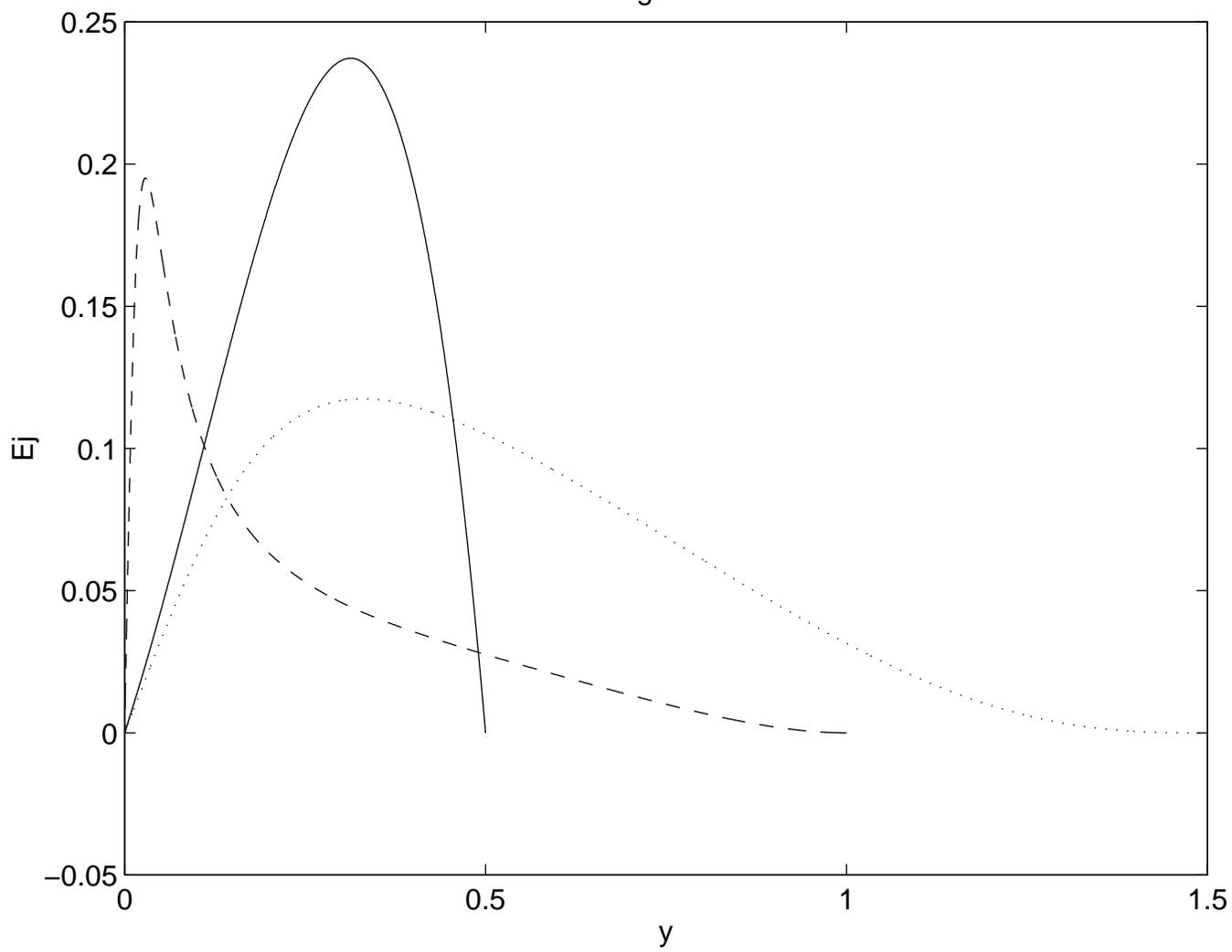
\end{document}